\documentstyle[multicol,aps,epsfig]{revtex}
\begin{document}
\newcommand{\be}{\begin{equation}}
\newcommand{\ee}{\end{equation}}
\title{Large extinctions in an evolutionary model: \\
The role of innovation and keystone species } 
\maketitle 

\vspace{0.2truecm}
\centerline{Sanjay Jain$^{*,\dag,\ddag,\P}$ and 
Sandeep Krishna$^{*,\P}$}
\centerline{$^*${\it Centre for Theoretical Studies, 
Indian Institute of Science, Bangalore 560 012, India}} 
\centerline{$^{\dag}${\it Santa Fe Institute, 1399 Hyde Park Road, Santa Fe, NM 87501, USA}}
\centerline{$^{\ddag}${\it Jawaharlal Nehru Centre for Advanced Scientific Research,
Bangalore 560 064, India}}
\centerline{$^{\P}$ Emails: jain@cts.iisc.ernet.in, sandeep@physics.iisc.ernet.in}
\begin{abstract}
{
The causes of major and rapid transitions observed in biological
macroevolution as well as in the evolution of social systems are a
subject of much debate. Here we identify the proximate causes of
crashes and recoveries that arise dynamically in a model system
in which populations of (molecular) species co-evolve with their
network of chemical interactions. Crashes are events that involve
the rapid extinction of many species and recoveries the assimilation
of new ones. These are analyzed and classified 
in terms of the structural properties
of the network. We find that in the absence of large external 
perturbation, `innovation' is a major cause of large extinctions
and the prime cause of recoveries. Another major cause of crashes
is the extinction of a `keystone species'.
Different classes of causes produce crashes of different
characteristic sizes.}
\end{abstract}
     
\begin{multicols}{2}

Major transitions in 
biological and social systems are attributed primarily to
some combination of external perturbation, selection, novelty, and
a complex internal dynamics and structure of the system
\cite{Maynard-Smith,Raup,GE,Kauffman,BS,Glen,CD,Padgett}.
While empirical studies attempt to identify proximate causes 
for individual events (which requires stupendous effort because
of the complexity of the systems and the difficulty of obtaining data)
most modeling efforts have tended to focus on the statistics of
the events taken as a whole (see, e.g., \cite{NP,Bouchaud}
for reviews of models in macroevolution and finance, respectively).
Here we discuss an evolutionary model that permits a precise dissection
of the (often multiple) causes of individual events.
While the model is abstract, highly simplified and 
motivated by {\it chemical} evolution, the structures
and processes that arise in it
seem to have the flavour of phenomena
in biological and social evolution. 
These include  the appearance of a `core' and `periphery' in the
network structure of the system, a shifting balance between
cooperative and competitive processes as structures evolve,
`core-shifts', history dependence, 
`keystone species', `innovations' that are `core-transforming'
or `dormant', and others. Their precise mathematical 
formulation 
and analysis of their role in major transitions 
in this simplified context may help
in constructing and analyzing more realistic models. 

\vspace*{0.5cm}
\noindent
{\bf The Model} \\
The system is an idealized  
prebiotic pond containing a set of $s$ chemical species.
A given species $j$ can be a catalyst for the production of 
another species $i$ with some small {\it a priori} probability $p$.
Then the presence of $j$ in the pond causes the population of $i$ to
increase according to the rate equation for catalyzed chemical reactions.
The catalytic relationship is represented graphically by 
an arrow from node $j$ to node $i$ in a directed graph representing 
the chemical network. 
The graph is completely specified by its adjacency matrix
$C=(c_{ij})$, $i,j \in \{1,\ldots,s\}$. $c_{ij}$ is unity if there is a link from
node $j$ to node $i$, i.e., if species $j$ catalyzes the production
of species $i$,
and zero otherwise. Each species has
a population $y_i$ in the pond. $x_i\equiv y_i/\sum_j y_j$ is its relative population;
by definition $0 \leq x_i \leq 1$,
$\sum_{i=1}^s x_i=1$. 
On a certain timescale $T$ the relative populations of the species 
reach a steady state denoted ${\bf X}\equiv (X_1,\ldots,X_s)$ that 
depends on the catalytic network.
We imagine the pond to be subject to periodic external perturbations on a timescale
greater than $T$ in the form of tides, storms or floods. Such a perturbation
can flush out existing molecular species from the pond (the ones with
the least $X_i$ being the most likely to be eliminated) and bring in
new chemical species from the environment whose catalytic relationships
with existing species in the pond are quite different from the eliminated ones.
Then, after the perturbation, the populations in the pond will evolve to a new 
steady state, be subject to another perturbation, and so on.

The precise rules are as follows: 
{\em Initialization:}
Start with a random graph with $s$ nodes and `catalytic probability' $p$.
That is, for every ordered pair $(i,j)$ with $i \neq j$, $c_{ij}$ is
unity with probability $p$ and zero with probability $1-p$.
$c_{ii}\equiv 0$
for all $i$ (to forbid self-replicating species).
Fig. 1a is an
example.
Assign each $x_i$ a random number between 0 and 1 and uniformly rescale all
$x_i$ so that $\sum_{i=1}^s x_i=1$.
{\em Dynamics:} First, keeping $C$
fixed, evolve ${\bf x}$ from its initial condition according to
\be
\dot{x}_i = \sum_{j=1}^s c_{ij} x_j -  x_i \sum_{k,j=1}^s c_{kj} x_j
\label{xdot}
\ee
for a time $T$ large enough to reach its attractor. Denote $x_i(T) \equiv X_i$.
Eq. \ref{xdot} follows from the rate equation
$\dot{y}_i=\sum_j c_{ij}y_j-\phi y_i$ for the populations, which in turn
is an idealization of rate equations in a well
stirred chemical reactor.\footnote
{The rate equation $\dot{y}_i=k(1+\nu y_j)n_An_B-\phi y_i$
follows from the reaction scheme
$A + B\stackrel{j}{\rightarrow} i$,
where $A$ and $B$ are reactants with populations $n_A$ and $n_B$,
$j$ and $i$ are catalyst and product with populations $y_j$ and
$y_i$ respectively, and $\phi$ is a death rate or dilution flux
in the reactor ($k$ is the rate constant for the spontaneous reaction
and $\nu$ is the catalytic efficiency).
If $n_A,n_B$
are large and fixed and the spontaneous reaction is much slower then
the catalyzed reaction, this equation reduces to
$\dot{y}_i=cy_j-\phi y_i$, where $c$ is a constant.
A generalization of the latter equation is
$\dot{y}_i = \sum_{j=1}^{s}c_{ij}y_j-\phi y_i$ for the case where
species $i$ has multiple catalysts. 
Eq. \ref{xdot} follows from this by taking the time derivative 
of $x_i=y_i/ \sum_{j=1}^s y_j$. In the present model we make the idealization
that all catalytic strengths are equal. The second (quadratic)
term in Eq. \ref{xdot} is needed to preserve
the normalization of the $x_i$ under time evolution.
Note that it follows automatically from the nonlinear
relationship between $x_i$ and $y_i$ when the time derivative of
$x_i$ is taken.}
Find the set of nodes with the least $X_i$.
Second, pick a node (denoted $k$) from this set at random 
and remove this node from the graph along with all its links. 
Add a new node (also denoted $k$) to the graph which is connected randomly
to the existing nodes according to the same catalytic probability $p$.
Mathematically, this means that for every $i \neq k$,
$c_{ki}$ and $c_{ik}$ are reassigned to unity with probability $p$ and zero with
probability $1-p$, irrespective of the value they had earlier, and $c_{kk}=0$.
Set $x_k=x_0$ (a small constant),
perturb all other $x_i$ about their existing value $X_i$ by a small amount, and uniformly rescale all $x_i$
to preserve the normalization
$\sum_{i=1}^s x_i=1$. This procedure provides a new graph and a new
initial condition for ${\bf x}$. Now return to the first step of the
dynamics and iterate the procedure several 
times.\footnote{The attractor configuration
${\bf X}$ is determined in this paper by its algebraic properties discussed 
later, not by numerically integrating Eq. \ref{xdot}. Hence we are effectively 
taking $T=\infty$ which allows us to prove certain results analytically.
Preliminary simulations suggest that when Eq. \ref{xdot} is numerically
integrated for a fixed, but large, $T$, qualitatively similar results
are obtained, although one can expect that some of the sharp transitions
are now spread out over several graph update time steps.
}

This model, introduced in ref. 
\cite{JK1}, 
was inspired by the work in
refs. 
\cite{Dyson,FKP,BFF,FB,Kauffman,BS}. 
The removal of the 
least populated species implements selection 
\cite{BS},
and its replacement by another randomly connected species implements
the introduction of novelty into the system. 
The network 
character is shaped by the repeated application of these two incremental
external perturbations along with the internal population dynamics.

\vspace*{0.5cm}
\noindent
{\bf Three regimes of behaviour and transitions between them.} 
The system exhibits three regimes or phases of behaviour. 
This is illustrated in Fig. 2 which shows the number of populated species in the attractor, 
$s_1$, (those species for which $X_i>0$) vs. time ($n$, the number of
graph updates) for 
a run with $s=100$ and $p=0.0025$.
In the `random phase'
$s_1$ stays low with small fluctuations.
In the `growth
phase', $s_1$ typically rises exponentially with occasional drops.
Finally, in the `organized phase' 
$s_1$ stays close to $s$, the maximum value it can take.
The random and growth phases were discussed in \cite{JK1,JK2,JK3}.
As is evident from Fig. 2, the organized and the growth phases
exhibit occasional discontinuous transitions or `crashes' in which a number of
species suddenly go extinct (their $X_i$ become zero in a single time step).
At the end of a crash the system is in the random or growth phase. This is
followed by a recovery in which the system moves again towards the organized
phase. In \cite{JK4} it is shown that crashes are primarily `core-shifts',
a specific kind of change in the structure of the graphs (discussed below)
and recoveries are due to the growth of `autocatalytic sets'.
The main purpose of this paper is to elucidate the mechanisms which cause core-shifts.
Crashes and recoveries also occur in a related model with negative links and 
variable link strengths \cite{JK3}, where they are more difficult to study 
analytically.

\vspace*{0.5cm}
\noindent
{\bf Definitions and notation\\} 
{\it Autocatalytic set (ACS)}: 
An ACS is a subgraph each of whose nodes has at least one incoming link
from a node of the same subgraph. (By a subgraph we mean a subset of nodes
together with all their mutual links.) 
Thus an ACS contains a catalyst for each of its 
members 
\cite{Eigen,Kauffman2,Rossler}. 
In each of Figs. 1b-k, the subgraphs formed by the set of
all black nodes or all the black and grey nodes are ACSs. 
Figs. 1a and 1l do not have an ACS.
For any subgraph $A$, let $\lambda_1(A)$ be the largest eigenvalue of the adjacency
matrix of $A$. We  denote $\lambda_1(C)\equiv \lambda_1$.
It can be shown 
\cite{JK2}
that if the graph does not have an ACS, then $\lambda_1=0$, and if it does, then $\lambda_1 \geq 1$.
$\lambda_1$ therefore represents a topological property of the network.

\noindent
{\it Dominant ACS}: 
It can be shown that if $\lambda_1\ge 1$
the subgraph comprising the populated species ($X_i>0$) must be an ACS \cite{JK1,JK2},
which will be referred to as the dominant ACS. The dominant ACS
is uniquely determined by the graph and does not depend upon the
initial condition for ${\bf x}$ (except for special initial conditions
forming a set of measure zero, which we ignore).
The subgraph formed by the set of all black and grey nodes in each of Figs. 1b-k is the
dominant ACS for that graph.
In addition to its topological significance, $\lambda_1$ also has a 
dynamical interpretation as being the `population growth rate' of the dominant 
ACS.\footnote{
This follows from the fact
that the attractor configuration ${\bf X}$ is always an eigenvector of $C$ with eigenvalue
$\lambda_1$, i.e., $\sum_jc_{ij}X_j=\lambda_1 X_i$ \cite{JK1}. Thus,
when $\phi=0$, substituting $y_i\propto X_i$ in the population dynamics equation
$\dot{{\bf y}}=C{\bf y}$, one gets $\dot{{\bf y}}=\lambda_1{\bf y}$. }

\noindent
{\it Core} and {\it periphery} of the dominant ACS:
The core of the dominant ACS of a graph $C$ (sometimes also referred
to as the `core of $C$')
is the maximal subgraph, $Q$, 
from each of whose nodes all
nodes of the dominant ACS can be reached along some directed
path. The rest of the dominant ACS is its periphery. 
The subgraph of all black nodes in Figs. 1b-k constitutes the core of the 
dominant ACS in the graph.\footnote
{Sometimes the dominant ACS consists of two
or more disjoint subgraphs, as in Fig. 1i. Then the above definition
applies to each component separately. There exist other ACS structures 
for which this definition is not adequate, e.g., two disjoint 2-cycles
pointing to a single downstream node. Such structures arise rarely and
can be treated by a more general definition of core and periphery
without altering the main conclusions presented here.}
Every periphery node has an incoming path that originates from the
core, but no outgoing path that leads to the core. The periphery
can contain loops within itself (e.g., the 2-cycle between nodes
$36$ and $74$ in Fig. 1e). 
One can prove that $\lambda_1(Q)=\lambda_1(C)\equiv\lambda_1$.

A subgraph with 
more than one node containing a directed path from each of its nodes to each
of its other nodes is said to be {\it irreducible} \cite{Seneta}. An 
irreducible subgraph is always an ACS but the converse is not true. However,
the core (of each component of a dominant ACS) is an irreducible subgraph. 
Since $\lambda_1(A) \geq 1$ for any irreducible subgraph $A$, it follows that
the latter is a `self-sustaining' structure in the sense
that if no other links were present in the graph, the nodes of $A$
would still have nonzero $X_i$ by virtue of their mutual links.
$\lambda_1$ measures the `strength' of the core in two ways: one, its
intrinsic population growth rate, and two, its multiplicity 
of internal pathways. To see the latter, compare
the increasing and decreasing pattern of $\lambda_1$ between $n=2854$
and $5042$ in the inset of Fig. 2 with the sequence of Figs. 1b-f.
When the core (of every disjoint component of the dominant ACS)
has exactly one cycle (Figs. 1b,f-i,k), then $\lambda_1 = 1$, and
vice versa. Such a core is fragile due to the
absence of any redundancy in its internal pathways;
the removal of any link from such a core will cause the ACS
property (of that component) to disappear. 

\noindent
{\it Crash}: a crash is a graph update event, $n$, in which a significant number
(say $ > s/2$) of the species go extinct, i.e., $\Delta s_1(n)
\equiv s_1(n)-s_1(n-1) < -s/2$.

\noindent
{\it Core overlap}:
given any two graphs $C$ and $C'$ whose nodes are labeled,
the core overlap between them, denoted $Ov(C,C')$, is
the number of common links in the cores of $C$ and $C'$. 
If any one of them does not have an ACS, $Ov(C,C')\equiv 0$.

\noindent
{\it Core-shift}: A core-shift is a graph update event in which $Ov(C_{n-1},C_n)=0$, i.e,  
there is no overlap between the cores of the dominant ACS before and after the event
($C_n$ is the graph at time step $n$).

\vspace*{0.5cm}
\noindent
{\bf Keystone species.}
One can consider the impact of the hypothetical removal of any species $i$
from the graph, irrespective of whether $i$ is the least populated or not.
For example one can ask for the core of the graph $C-i$ that would result
if species $i$ (along with all its links) were removed from $C$.
A species $i$ will be referred to as a {\it keystone species} 
if $Ov(C,C-i)=0$. Thus a keystone species is one whose removal modifies the
organizational structure of the graph (as represented by its core) drastically. 
In the ecological literature a keystone species is regarded as one
whose elimination from the ecosystem would
cause a significant fraction of species in the ecosystem to go extinct
\cite{Paine,Pimm,JTM,SMo}. 
We will see that likewise in the present model the removal of keystone species
causes large crashes. (Indices of keystoneness based on 
$Ov(C,C-i)$ or on the change in $s_1$ caused by the removal of a species
can also be defined.) 
Note that if $\lambda_1=1$ and the ACS has a single
connected component (e.g., figs. 1b,f-h,k), every core node is necessarily
a keystone species since its removal would destroy the cycle that 
constitutes the core.

\vspace*{0.5cm}
\noindent {\bf Innovations.} In the present model the new species $k$
at each time step together with its
set of new links may be regarded as a `novelty'
introduced into the system. In the new attractor the new species may go
extinct, i.e., $X_k$ may be zero, or it may survive, i.e., $X_k$ is non-zero.
Let us define an {\it innovation} to be a novelty in which the 
relative population of the new species in the new attractor 
just after the novelty occurs is nonzero. 
This definition has the feature that an innovation always involves 
{\it new connections}. It does not use any exogenously defined notion 
of fitness. The only performance criterion it requires is that 
the new links should enable the new node to {\it survive} until the next
graph update. Even this minimal requirement has nontrivial consequences.  
For instance when the new species receives an incoming link from an 
existing dominant ACS, it typically has a nonzero population in the new
attractor. Each recovery process, which occurs due to the
expansion of the dominant ACS during the growth phase \cite{JK4},
is an accumulation of just such innovations.  

We will be interested in a special class of innovations in which
the novelty creates a new populated irreducible subgraph. Such
innovations create or add to a `self-sustaining structure' in the
graph, in the sense discussed earlier.
For instance the appearance of the first ACS at $n=2854$ (see Fig.
1b) is such an innovation.  There, species 90, which was a singleton before the
event, went extinct and was replaced by a new species 90 that had an incoming
and outgoing link to node 26.  The two formed an irreducible subgraph whose
$\lambda_1$ value was 1.  This innovation in fact triggered the self
organization of the network around this ACS.  By $n=3880$ (see Fig.  1c) the
core had grown to 18 nodes as a result of several events in which the new
species was an addition to the core.  Every event in which the core is
strengthened by the addition of a new node is also an innovation in which a new
populated irreducible subgraph is formed.  When such an
innovation arises, denote the new irreducible subgraph which includes the new
species as $N$ (or $N_n$ at time step $n$.)  $N$ or $N_n$ will stand for the
maximal irreducible subgraph of which the new species $k$ is a member.  It
follows from the Perron-Frobenius theorem \cite{Seneta} that such an innovation
necessarily increases the $\lambda_1$ value of some substructure in the graph,
i.e., $\lambda_1(N)>\lambda_1(N-k)$. $N_n$ becomes the new core of the 
graph if it is `stronger' than the old core. More precisely, one can show
that $N_n$ will become the new
core of the graph, replacing the old core $Q_{n-1}$, whenever
any one of the following two conditions hold: \\
(a) $\lambda_1(N_n)>\lambda_1(Q_n')$ or, \\
(b)  
$\lambda_1(N_n)=\lambda_1(Q_n')$ and $N_n$ is `downstream' 
of $Q_n'$.\footnote{We use the notation $C_n' \equiv C_{n-1} - k$ for 
the graph of 
$s-1$ nodes just before the novelty at time step $n$ is brought in (and just
after the least populated species $k$ is removed from $C_{n-1}$). 
$Q_n'$ stands for the core of $C_n'$. A subgraph $A$ is `downstream' of another 
subgraph $B$ if there exists
a directed path from some node of $B$ to a node of $A$ but 
none from any node of $A$ to a node of $B$.} \\ 
Such an innovation may be referred to as a {\it core-transforming innovation}.
If $Q_{n-1} \subset N_n$, such an innovation enlarges the existing core.
However, if $Q_{n-1}$ and $N_n$ are disjoint, we get a core-shift.

In the present model, the
generation of novelty does not depend upon existing structure
since the links of a new node are chosen randomly from a fixed probability 
distribution: novelty is `noise' in this model. 
(Variants of the model that depart from this are easily constructed.)
However, whether the novelty constitutes an innovation is `context dependent' 
(i.e., dependent upon the structure of the existing network).
Also, the short and long term impact of an innovation depends upon
the (historical evolution of the) `context', as will be seen below.

\vspace{0.5cm} \noindent {\bf Classification of core-shifts} \\
In a set of runs with $s=100, p=0.0025$ totaling 1.55 million iterations we observed
701 crashes. 
612 of these were core-shifts
\cite{JK4}.
Fig. 3 differentiates between the 612 core-shifts we observed. 
They fall into three categories: (i) complete
crashes
($136$ events), 
(ii) takeovers by core-transforming innovations
($241$ events),
and (iii) takeovers by dormant innovations
($235$ events).

\vspace*{0.5cm}
\noindent
{\bf Complete crashes.} 
A complete crash is an event in which an ACS exists before but not after the graph update.
Therefore for a complete crash at time step $n$, $\lambda_1(C_{n-1})>0$ and $\lambda_1(C_n) = 0$.  
These events take the system back to the random phase. 
For example at $n=8232$, in Fig. 1k, node $54$ is one of the least
populated species and is hit at $n=8233$. It is replaced by a new species 
that has a single outgoing link to node $50$
and no incoming link, 
resulting in the complete disruption of the ACS. 
It is evident that complete crashes must always be caused by the
elimination of a keystone species.
Furthermore, Fig. 3  shows that 
$\lambda_1(C_{n-1})=1$ for every complete crash observed in the runs. Hence
the core of the ACS is a single cycle when the event occurs and the
species removed is a member of that cycle. 

\vspace*{0.5cm}
\noindent
{\bf Takeovers by core-transforming innovations.} 
An example of a takeover by a core-transforming innovation is given in 
Figs. 1g,h. At $n=6061$ the core was a single loop
comprising nodes $36$ and $74$. Node $60$ was replaced by a new species at $n=6062$.
The new node $60$ created an innovation at $n=6062$, with $N_{6062}$ being the cycle 
comprising nodes $60, 21, 41, 19$ and $73$, downstream from 
the old core. 
The graph at $n=6062$ has one cycle feeding into a second cycle that
is downstream from it. (This is an example of 
condition (b) for a core-transforming innovation.)
The population attractor of such a graph has the property that
the $X_i$ of the nodes in the upstream cycle are all zero
and only the nodes of the second cycle (as well as nodes further
downstream from it) have nonzero $X_i$. 
Thus when the above innovation arises, the new cycle becomes the
new core and all nodes that are not downstream
from it get $X_i=0$, resulting in a large drop in $s_1$ from 89 to 32. 
For all such events in Fig. 3, $\lambda_1(Q_n')=\lambda_1(C_{n-1})$ since
$k$ happened not to be a core node of $C_{n-1}$. 
Thus these core-shifts satisfy
$\lambda_1(C_n)=\lambda_1(N_n)\ge\lambda_1(Q_n')=\lambda_1(C_{n-1})\ge 1$ in 
Fig. 3.

\vspace*{0.5cm}
\noindent
{\bf Takeovers by dormant innovations.}
Figs. 1e,f show an example of a takeover by a dormant innovation.
At $n=5041$, the core has $\lambda_1=1.24$ and there is a cycle 
comprising nodes $36$ and $74$ in its periphery. Node $85$ is hit which 
results in a cycle ($26$ and $90$) feeding into another cycle ($36$ 
and $74$) at $n=5042$.
Thus at $n=5042$, for the same reason as in the previous paragraph,
$36$ and $74$ form the new core with only one other
downstream node, $11$, being populated. All other nodes become depopulated
resulting in a drop in $s_1$ by $97$. Such a core-shift is the 
result of an innovation that arose earlier (the cycle between
$36$ and $74$ arose at $n=4695$)
but lay dormant downstream of the
existing core until one of the keystone species of the latter (node
$85$) was hit and made it weak (i.e., reduced its $\lambda_1$ to 
a value less than or equal to the $\lambda_1$ of the downstream 
innovation). 
A dormant innovation can takeover as the new core
only following a keystone extinction which weakens the old core.
In such an event the new core necessarily has a lower (but nonzero)
$\lambda_1$ than the old core, i.e., $\lambda_1(C_{n-1})>\lambda_1(C_n)\ge 1$.
Note that at $n=5041$ if the downstream cycle between $36$ and $74$
were absent, $85$ would {\it not} be a keystone species by our 
definition, since its removal would still leave part of the
core intact (nodes $26$ and $90$). $85$ becomes
keystone, and the core of which it is a part becomes fragile
and susceptible to a core-shift,
{\it because} a self-sustaining innovation has occurred in 
the distant periphery. 

If we consider the set of all drops in $s_1$, large or small, where an
ACS exists before the event (there are 126454 such events in the 
above mentioned runs) we find that the number of complete crashes remains
the same, 136 (the mean size of the drop in such events is 
$\overline{|{\Delta s_1}|} = 98.2$ with a standard deviation $\sigma = 1.2$),
while core-shifts caused by dormant innovations go up to 359
(with $\overline{|{\Delta s_1}|} = 62.2$, $\sigma = 25.9$) and those
due to core-transforming innovations to 524 ($\overline{|{\Delta s_1}|} = 48.2$,
$\sigma = 25.6$). The rest of the events consist of 9851  `partial core-shifts'
(in which the core changes, but $Ov(C_{n-1},C_n) \neq 0$; in this category
$\overline{|{\Delta s_1}|} = 2.18$, $\sigma = 7.42$), 
and 115584 events where there was no change in the core
but the periphery is affected
($\overline{|{\Delta s_1}|} = 1.05$, $\sigma = 0.99$). 
Thus different classes of proximate causes of drops 
arise dynamically with different frequencies and typically produce
events in different size ranges. The ranges, however, overlap and some 
distributions have
fat tails (e.g., of the 701 crashes with $|{\Delta s_1}| > 50$, there 
are 79 and 10 events, respectively, in the last two categories). Detailed
distributions and their dependence on $s,p$ are open questions
(but see \cite{JK4}).

\vspace*{0.5cm}
\noindent
{\bf Discussion} \\
The present model exhibits mechanisms by which innovations 
can play a major role in crashes and recoveries in a
complex system.\footnote{Analogues of innovations
and core-shifts appear to be playing an important role
in another related but quite different model \cite{CRA} where
rapid transitions are observed.}
It provides a mathematical example of 
`creative destruction' \cite{Schumpeter} at work in causing large upheavals.
It distinguishes two processes involving 
innovations, both having analogues in the real world. 
One is exemplified by the 
appearance of the automobile
which made the horse drawn carriage and its ancillary industries
obsolete. This is like the example of the core-transforming innovation
shown in Figs. 1g,h where the graph update 
produced a self-sustaining structure that was more vibrant 
than the existing core within the context of the present organization. This structure became
the new core, rendering many nodes drawing sustenance from
the old core dysfunctional. 
The subsequent development of  
other industries dependent on the automobile mirrors the growth of
the ACS around the new core. 
The second process is exemplified by the emergence of
the body plans of several phyla which are dominant today.
It is believed that while these body plans originated 
in the Cambrian era more than 520 million years ago \cite{VJE},
the organisms with these body plans played no major role till
about 250 million years ago. They started flourishing only  
when the Permian extinction depleted the other species that
were dominant till that time 
\cite{Erwin}.
This is similar to the events shown in Figs. 1e,f where an earlier innovation
had lain dormant for a while without
disturbing the existing core, but when the latter became sufficiently
weak, took over as the new core and flourished. 

Recently there has been substantial progress in graph theoretic analyses
of complex systems and in particular `small-world' \cite{WS} and
`scale-free' \cite{BA} properties have been found for several real
networks (for reviews and references see \cite{Watts,AB}). It may be
interesting to study whether certain classes of real networks also have
some kind of a `core-periphery' structure, in view of its possible dynamical
significance.

\vspace*{0.5cm}
\noindent
{\bf Acknowledgements.}
S.J. acknowledges the Associateship of Abdus Salam International
Centre for Theoretical Physics, Trieste as well as the hospitality of the Max Planck
Institute for Mathematics in the Sciences, Leipzig, and Jesus College,
Cambridge.
This work was supported in part by a grant from
the Department of Science and Technology, Government of India.

\vspace*{0.5cm}
\noindent {\bf Figure legends\\} 

\noindent
{\bf Figure 1.} The structure of the evolving graph at various time instants
for a run with $s=100, p=0.0025$. Node numbers $i$ from $1$ to $100$
are shown in the circles representing the nodes. Nodes with 
zero relative population in the attractor configuration 
for the graph ($X_i=0$) are shown as
white circles; the rest ($s_1$ in number) have nonzero $X_i$.
In the graphs where an autocatalytic set (ACS) exists, black circles correspond to nodes in the `core' of the ACS,
and grey to the `periphery', defined in the text.
{\bf (a)} $n=1$, the initial random
graph, {\bf (b)} $n=2854$, where the first ACS, a 2-cycle between
nodes $26$ and $90$, appeared, {\bf (c)} $n=3880$, the beginning of the
organized phase when the ACS first
spanned the entire graph, {\bf (d)} $n=4448$, when the core reached 
a peak in the number of loops it contained, 
{\bf (e)} $n=5041$, just before a `core-shift', {\bf (f)} $n=5042$, just 
after the core-shift caused by a `keystone' extinction in the
presence of a `dormant innovation', 
{\bf (g)} $n=6061$, just before another core-shift, 
{\bf (h)} $n=6062$, just after the core-shift caused by a `core-transforming innovation',
{\bf (i)} $n=6070$, when the old core stages a come back as a disconnected
component after node $32$ becomes a singleton,
{\bf (j)} $n=6212$, when the new core strengthens itself and 
depopulates the recently resurgent old core,
{\bf (k)} $n=8232$, just before the first `complete crash',
{\bf (l)} $n=10000$, between the first `complete crash' and the subsequent `recovery'.\\

\noindent
{\bf Figure 2.} The number of populated species, $s_1$ (continuous line), and
the largest eigenvalue of $C$ (whose significance is discussed later in the text),
$\lambda_1$ (dotted line), versus time, 
$n$ for a run with $s=100$ and $p=0.0025$. The $\lambda_1$ values shown are $100$ times the actual $\lambda_1$ value. 
The first $10000$ time steps are enlarged in the
inset. The run is the same for which the
graph snapshots are shown in Fig. 1. The impact of the events described
in Fig. 1 is clearly visible in this curve. E.g., at $n=2854$ $\lambda_1$ jumps
from zero to one and $s_1$
exhibits the first sustained upward movement, at $n=3880$ $s_1$ hits its
maximum value, $100$, and then fluctuates mainly between $99$ and 
$100$, and at $n=4448$ $\lambda_1$ reaches a local maximum.
$s_1$ drops from $100$ to $3$ as a result of the `core
shift' at $n=5042$, and from $89$ to $32$ in the core-shift at 6062.
At $n=6070$ a large recovery event is seen as the old core and 
the still intact part of its periphery get repopulated, only to be 
trounced again at $n=6212$ when the new core strengthens itself to
a $\lambda_1$ value greater than $1$. 
At $n=8233$ $s_1$ crashes from $100$ to $2$ when the ACS is
completely destroyed and $\lambda_1$ drops from one to zero.\\

\noindent
{\bf Figure 3.} Frequency, $f$, of the $612$ core-shifts observed in a set of runs with $s=100$ and 
$p=0.0025$ vs. the
$\lambda_1$ values before, $\lambda_1(C_{n-1})$, and after, $\lambda_1(C_n)$, the core-shift. 
Complete crashes 
(black; $\lambda_1(C_{n-1})=1$, $\lambda_1(C_n)=0$),
takeovers by core-transforming innovations (blue; $\lambda_1(C_n)\ge\lambda_1(C_{n-1})\ge 1$) 
and takeovers by dormant innovations (red; $\lambda_1(C_{n-1})>\lambda_1(C_n)\ge 1$) are
distinguished. Numbers alongside vertical lines represent the corresponding $f$ value.

\end{multicols}
\pagebreak
\vspace{-3.5cm}
\begin{multicols}{3}
\vspace{-2cm}
\begin{figure}
\epsfysize=6cm
\noindent
{\bf (a) n=1\\}

\vspace{-0.8cm}
\noindent
\epsfbox{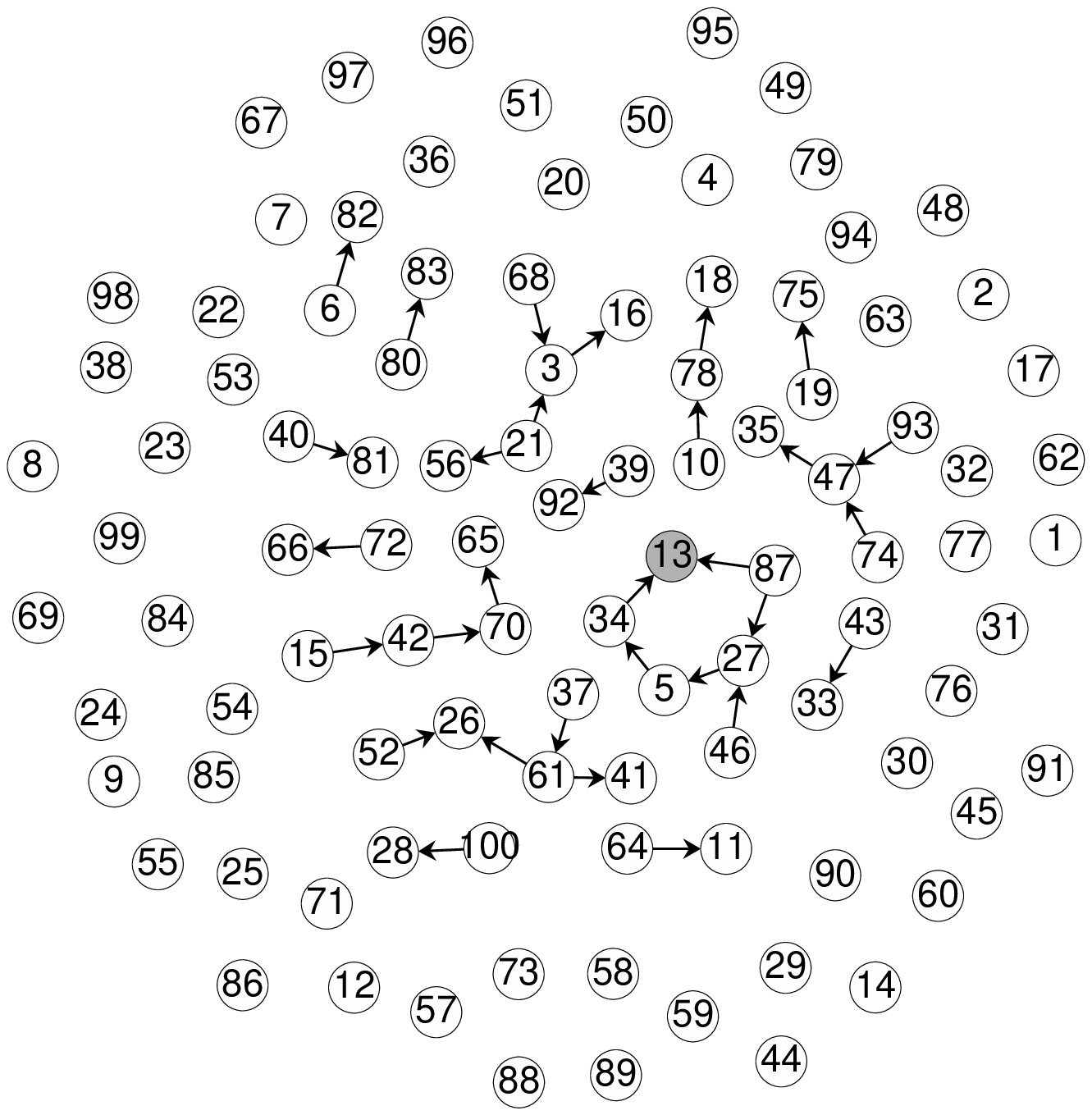}
\end{figure}

\vspace{-1cm}
\begin{figure}
\epsfysize=6cm
\noindent
{\bf (d) n=4448\\}

\vspace{-1.1cm}
\noindent
\epsfbox{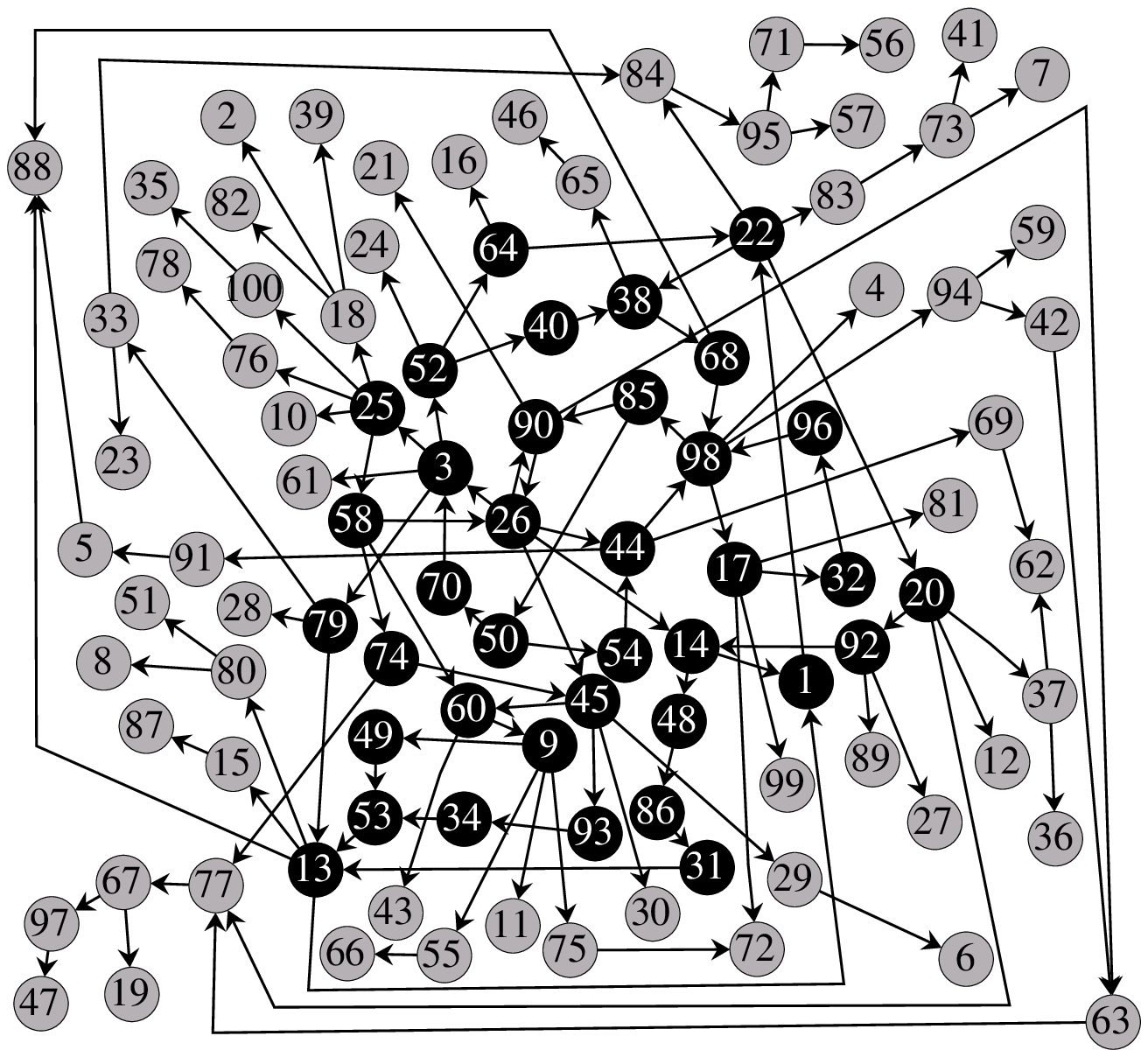}
\end{figure}

\vspace{-1cm}
\begin{figure}
\epsfysize=6cm
\noindent
{\bf (g) n=6061\\}

\vspace{-0.8cm}
\noindent
\epsfbox{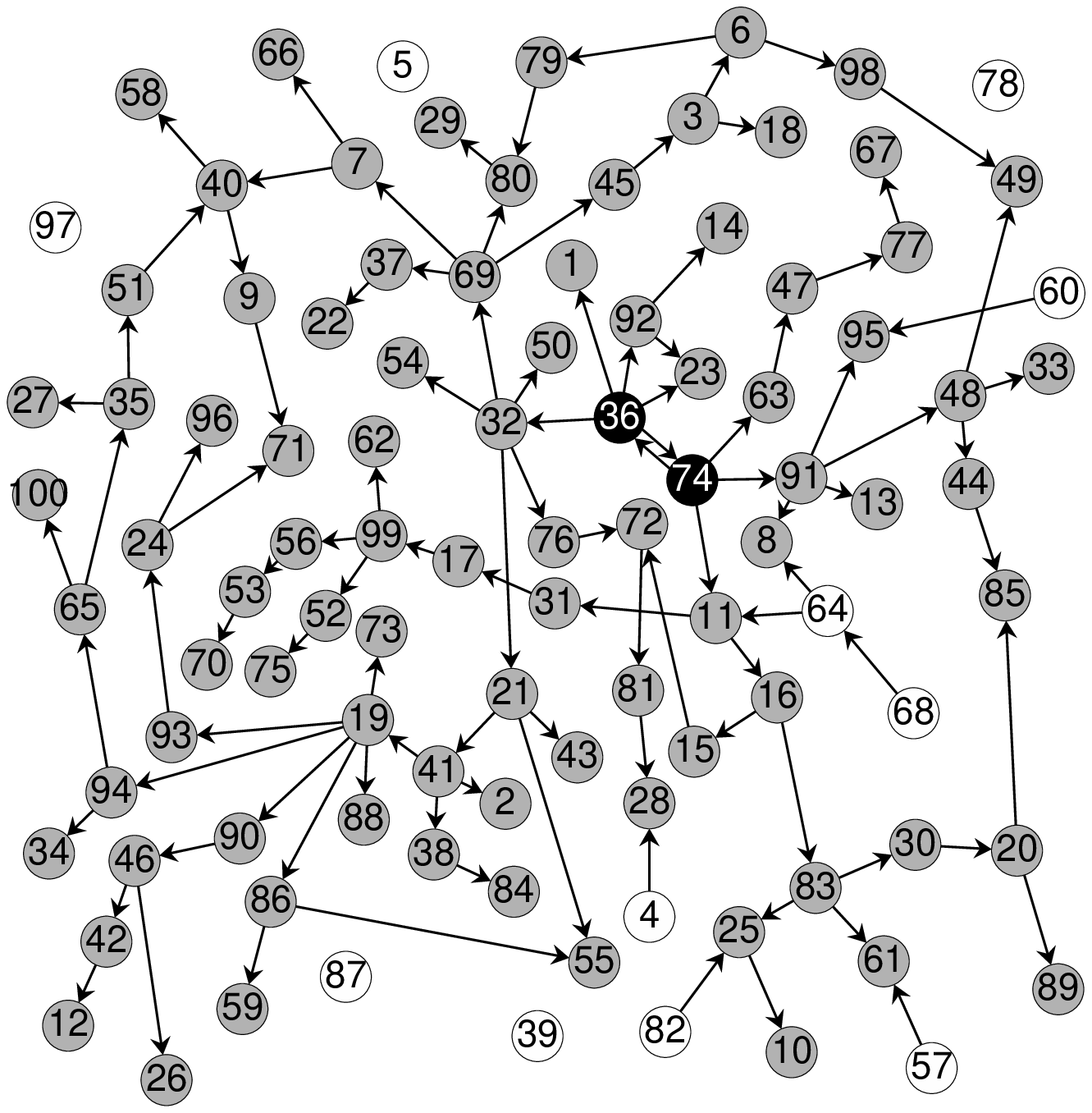}
\end{figure}

\vspace{-1cm}
\begin{figure}
\epsfysize=6cm
\noindent
{\bf (j) n=6212\\}

\vspace{-0.5cm}
\noindent
\epsfbox{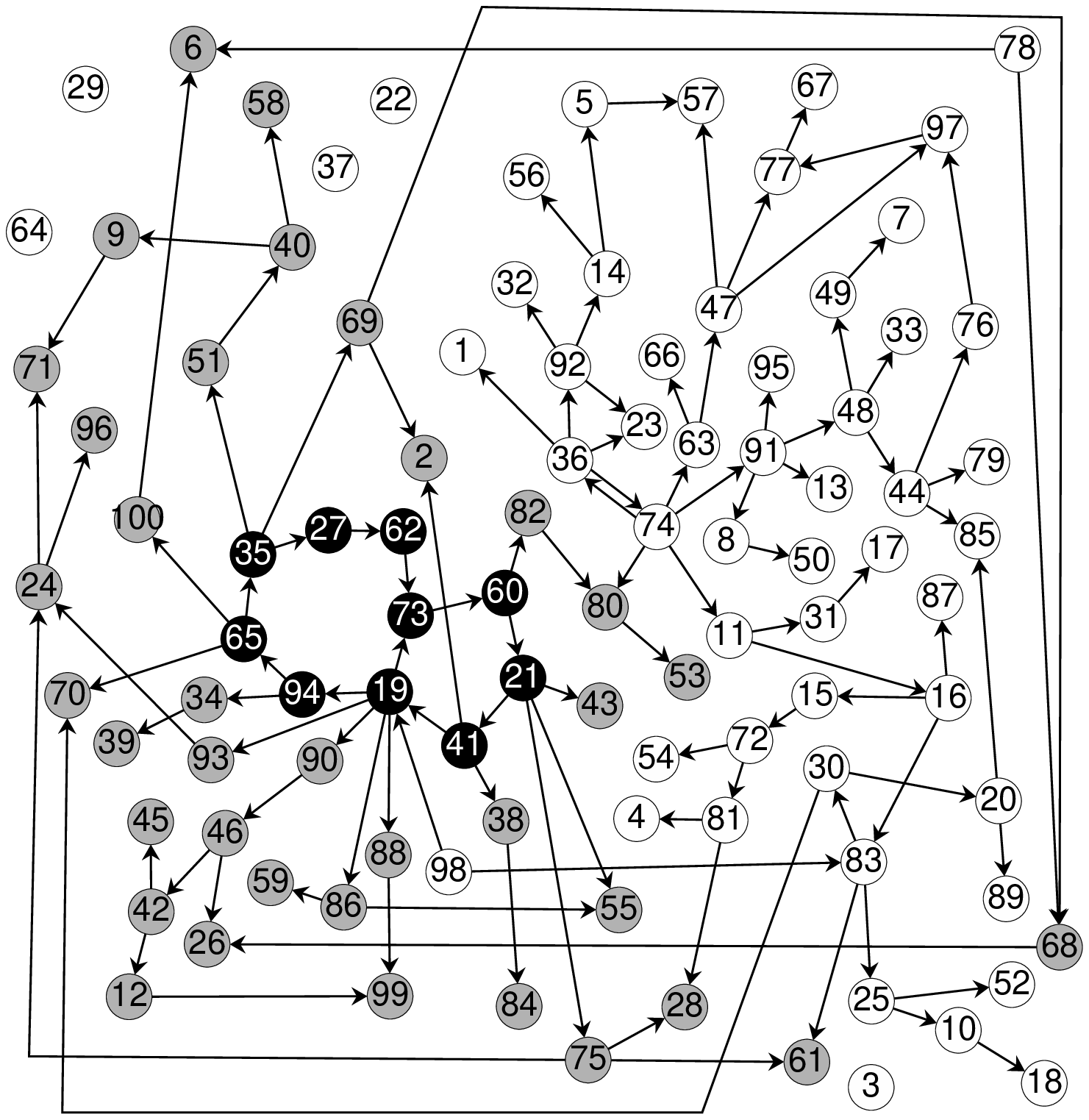}
\end{figure}

\begin{figure}
\epsfysize=6cm
\noindent
{\bf (b) n=2854\\}

\vspace{-0.8cm}
\noindent
\epsfbox{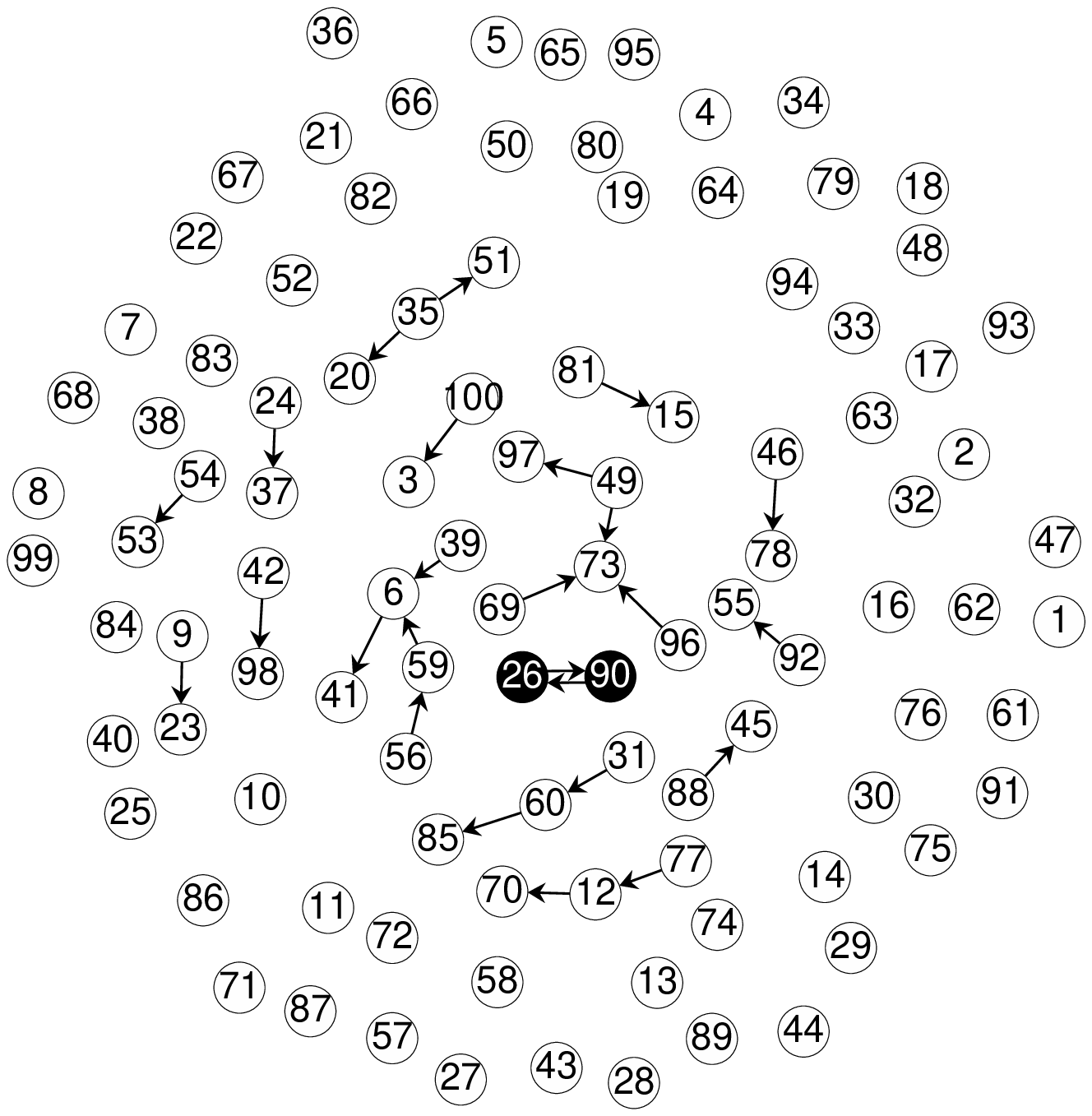}
\end{figure}

\vspace{-1cm}
\begin{figure}
\epsfysize=6cm
\noindent
{\bf (e) n=5041\\}

\vspace{-0.8cm}
\noindent
\epsfbox{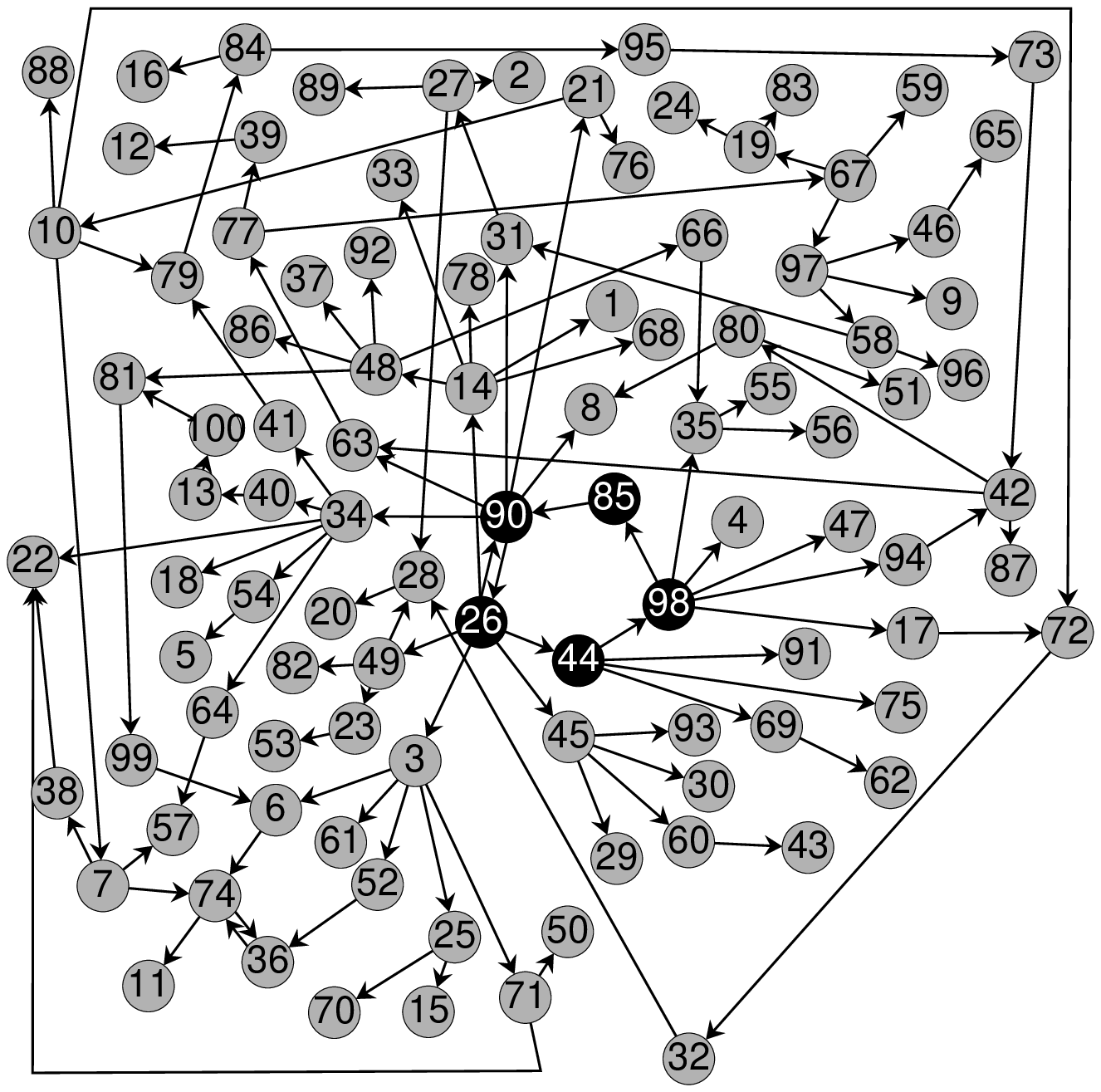}
\end{figure}

\vspace{-1cm}
\begin{figure}
\epsfysize=6cm
\noindent
{\bf (h) n=6062\\}

\vspace{-0.9cm}
\noindent
\epsfbox{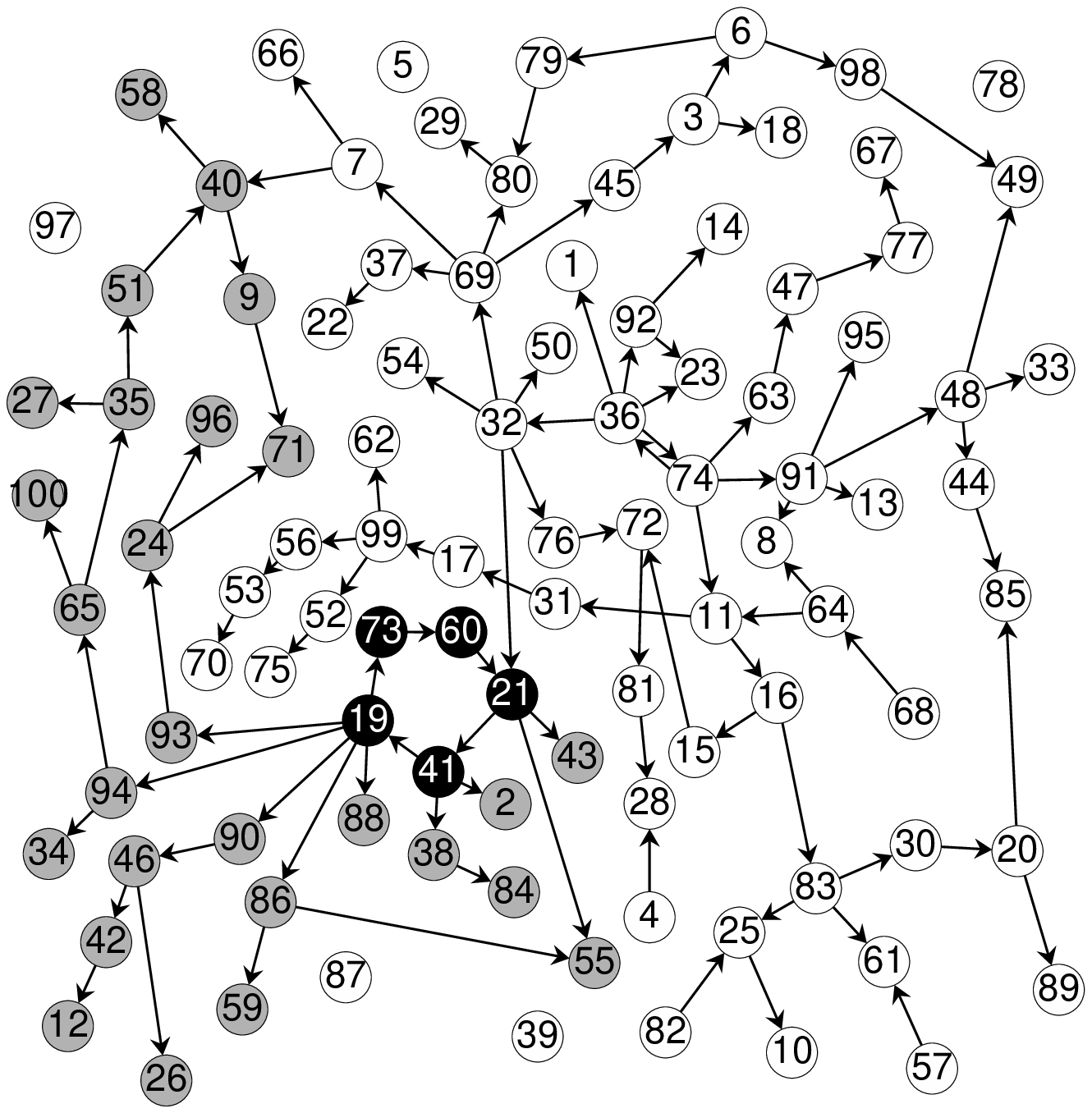}
\end{figure}

\vspace{-1cm}
\begin{figure}
\epsfysize=6cm
\noindent
{\bf (k) n=8232\\}

\vspace{-1.2cm}
\noindent
\epsfbox{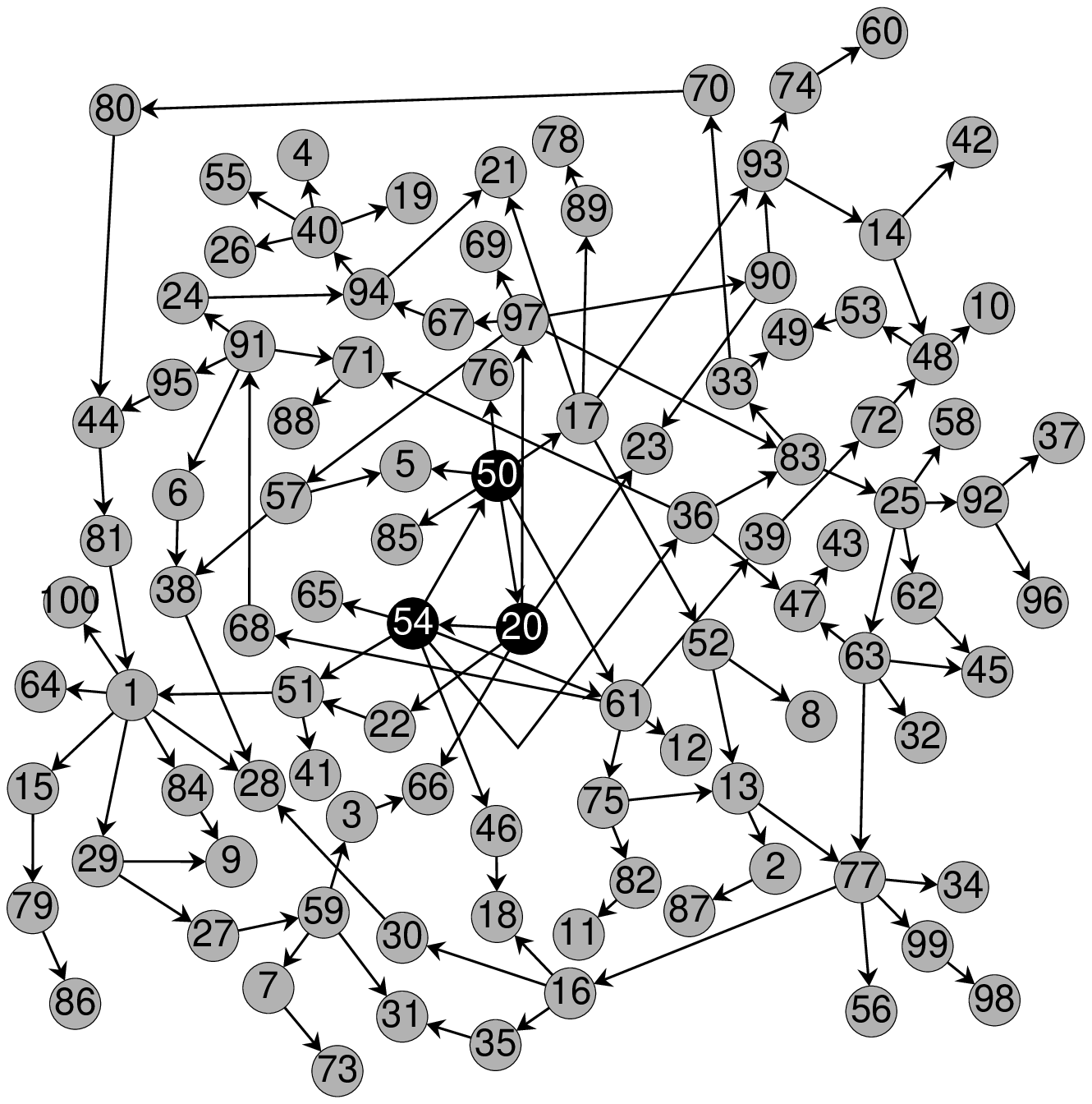}
{\bf \centerline{Figure 1.}}
\end{figure}

\vspace{-2.5cm}
\begin{figure}
\epsfysize=6cm
\noindent
{\bf (c) n=3880\\}

\vspace{-0.8cm}
\noindent
\epsfbox{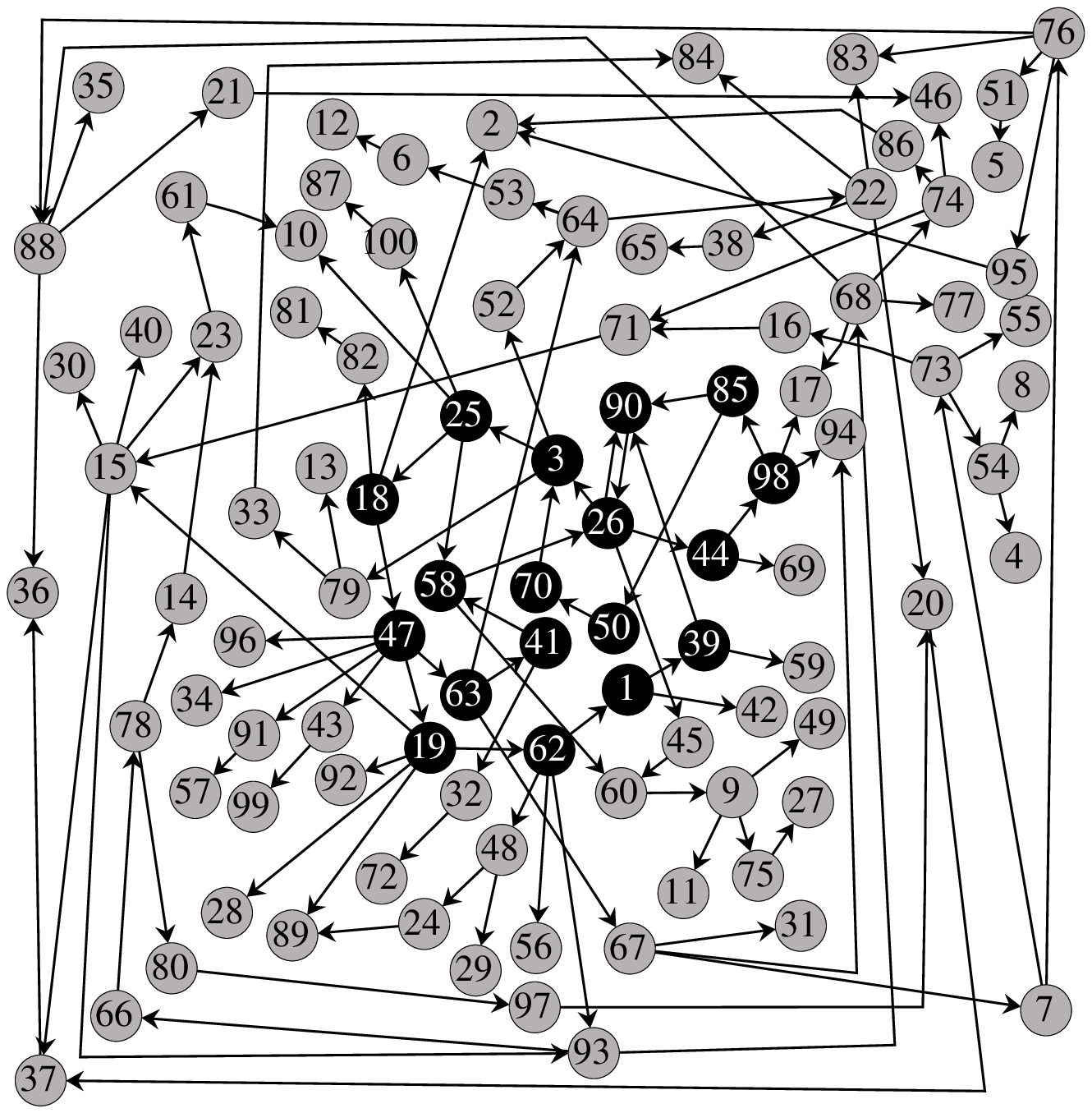}
\end{figure}

\vspace{-1.2cm}
\begin{figure}
\epsfysize=6cm
\noindent
{\bf (f) n=5042\\}

\vspace{-0.8cm}
\noindent
\epsfbox{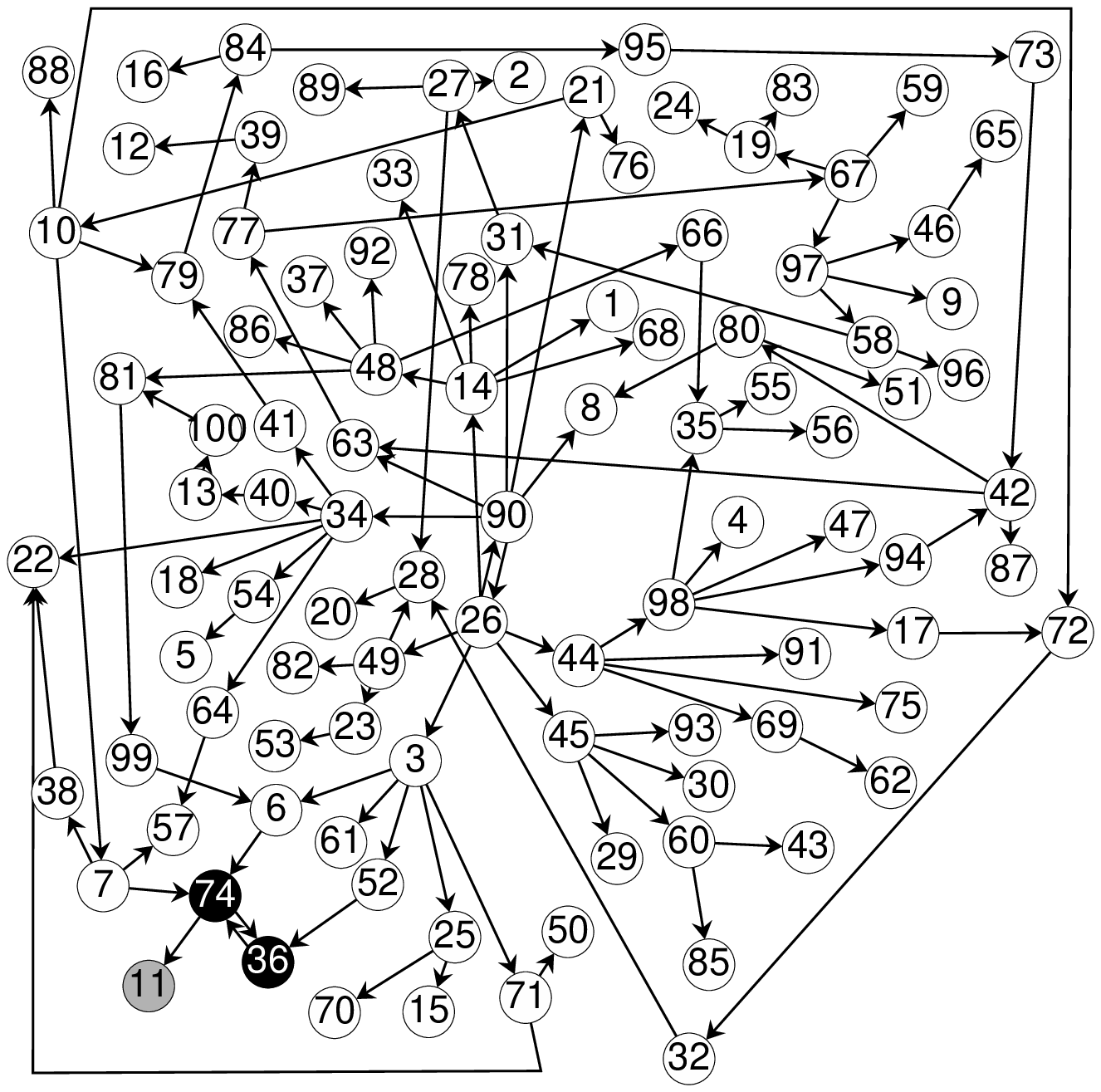}
\end{figure}

\vspace{-1.2cm}
\begin{figure}
\epsfysize=6cm
\noindent
{\bf (i) n=6070\\}

\vspace{-0.8cm}
\noindent
\epsfbox{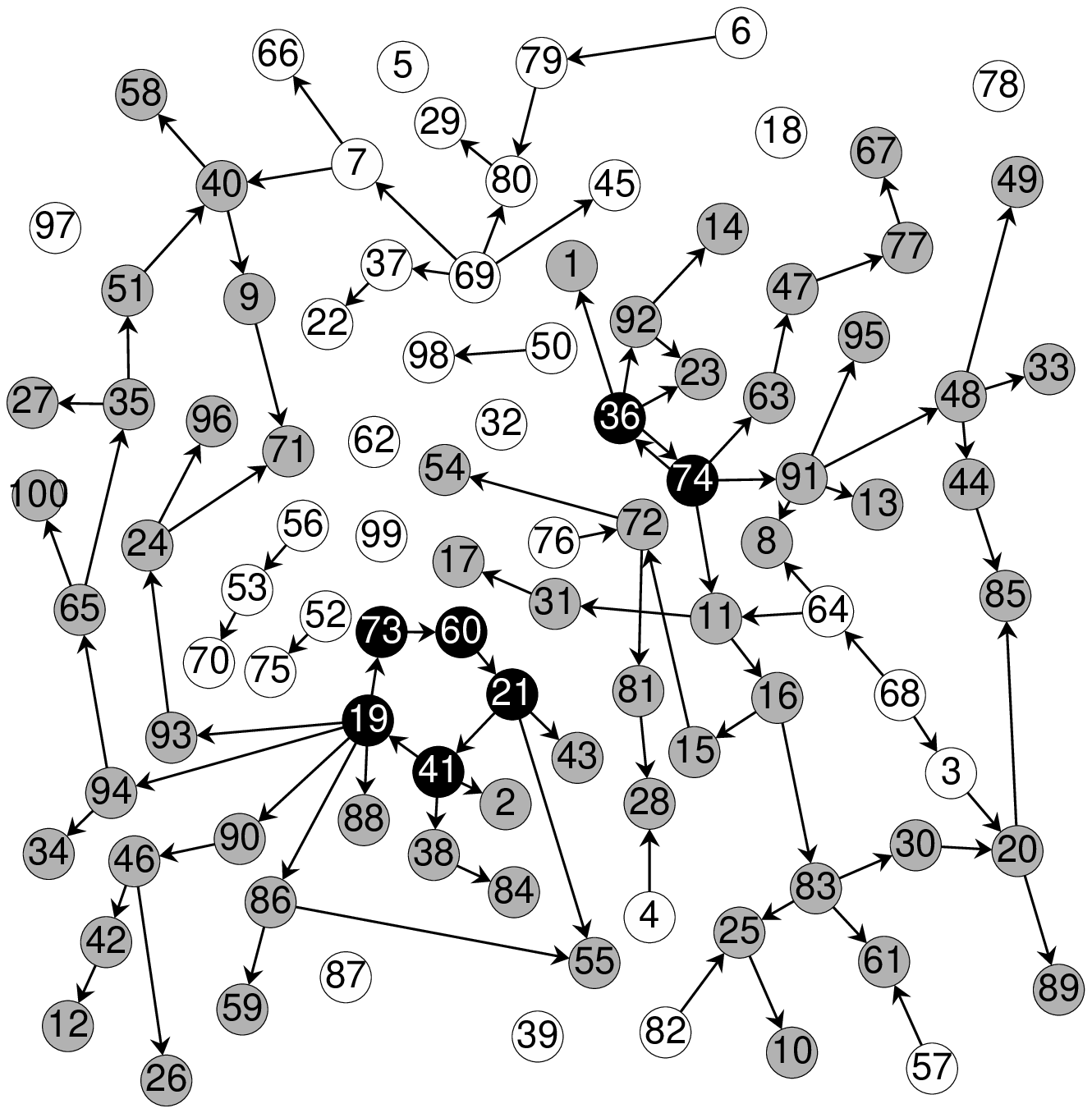}
\end{figure}

\vspace{-1.1cm}
\begin{figure}
\epsfysize=6cm
\noindent
{\bf (l) n=10000\\}

\vspace{-1.1cm}
\noindent
\epsfbox{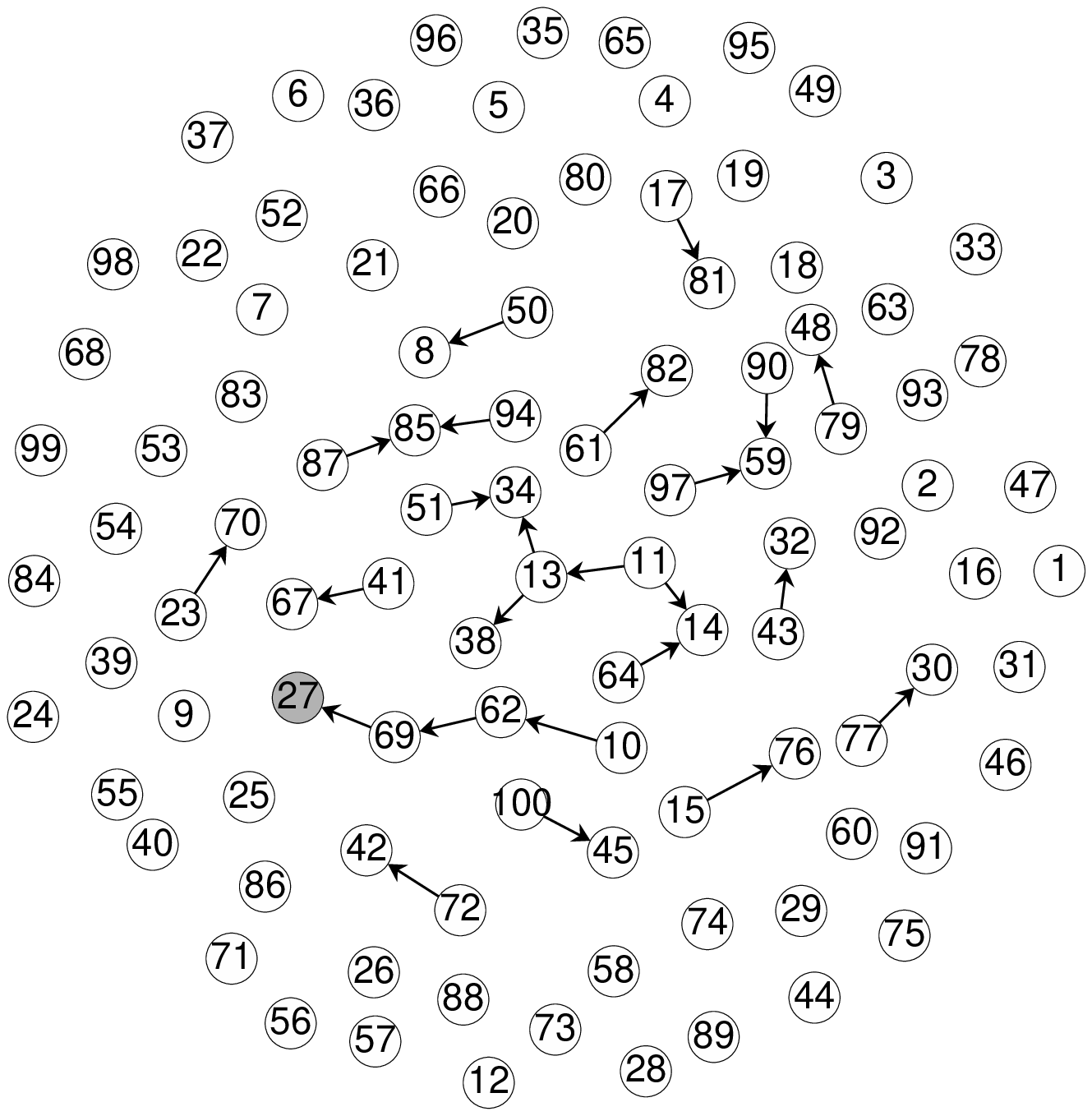}
\end{figure}

\end{multicols}


\begin{multicols}{2}

\begin{figure}
\epsfxsize=8cm
\noindent
\epsfbox{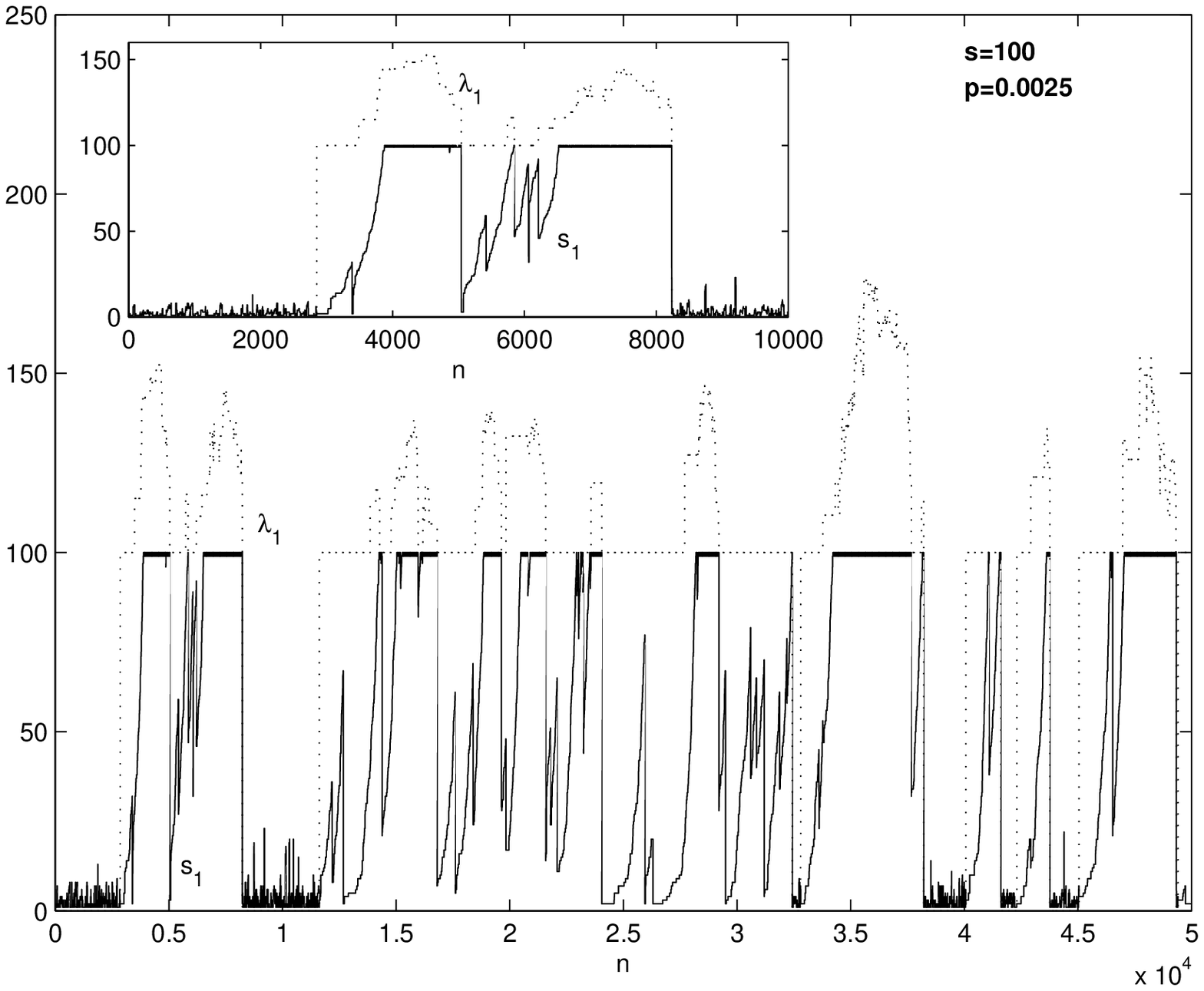}
{\bf Figure 2.} 
\end{figure}

\vspace*{1cm}
\begin{figure}
\epsfxsize=8cm
\noindent
\epsfbox{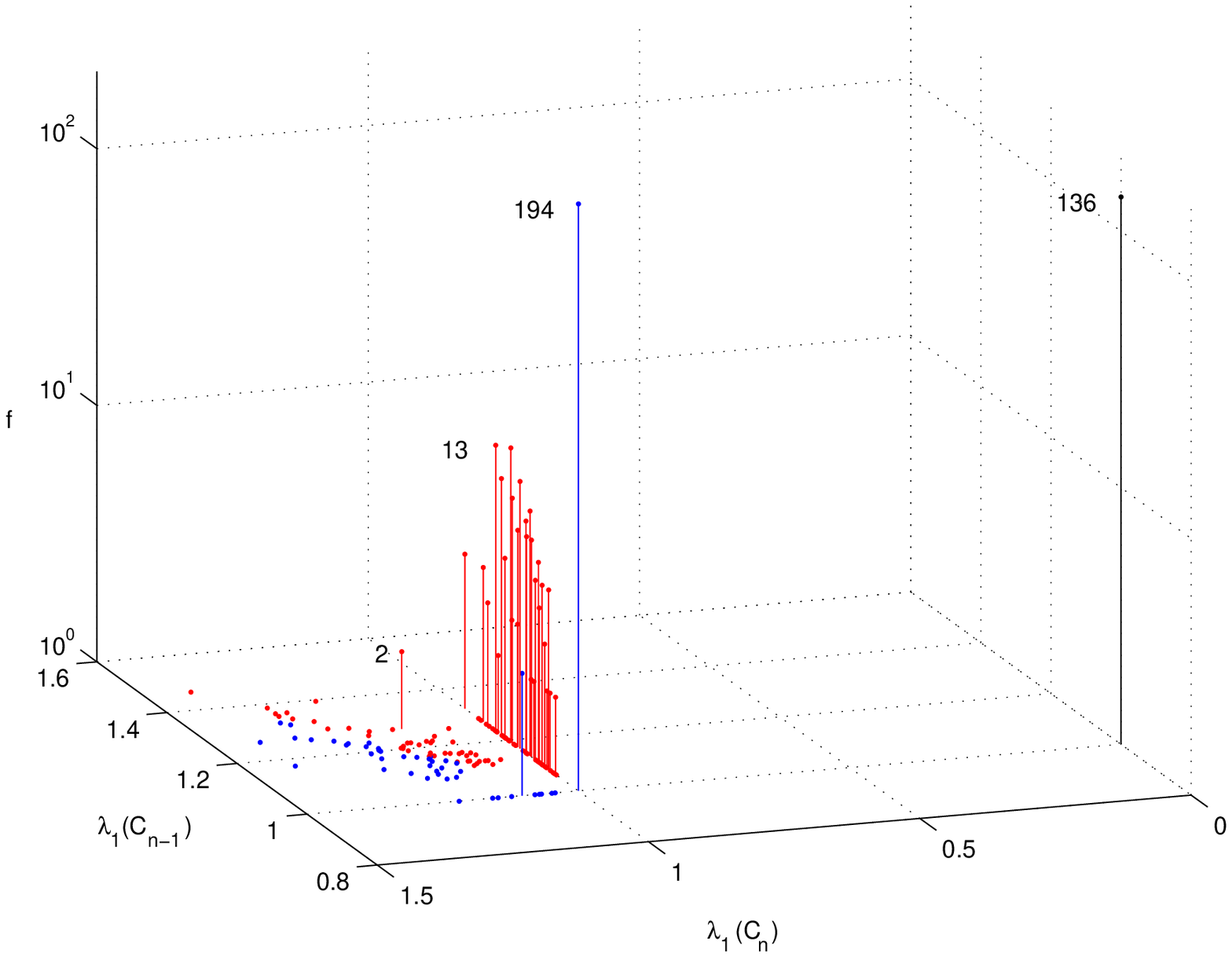}
\end{figure}

\noindent
\vspace*{-0.5cm}
{\bf Figure 3.} 

\end{multicols}

\begin{thebibliography}{99}
\bibitem{Maynard-Smith}{Maynard-Smith, J. (1989) {\it Phil. Trans. R. Soc.}
{\bf B325} 241-252.}

\bibitem{Raup}{Raup, D. M. (1991) {\it Extinction: Bad genes or bad luck?} (Norton,
New York).}

\bibitem{GE}{Gould, S. J. \& Eldredge, N. (1993) {\it Nature} {\bf 366}
223-227.}

\bibitem{Kauffman}
{Kauffman, S. A. (1993)
{\it The Origins of Order}
(Oxford Univ. Press).}

\bibitem{BS}
{Bak, P. \& Sneppen, K. (1993)
{\it Phys. Rev. Lett.} {\bf 71}, 4083-4086.}

\bibitem{Glen}{Glen, W. (editor) (1994) {\it The mass extinction debates} (Stanford 
Univ. Press, Stanford).} 

\bibitem{CD}{Carlson, J. M. \& Doyle, J. (1999) {\it Phys. Rev.} {\bf E60}
1412-1427.}

\bibitem{Padgett}{Padgett, J. (2001) in
{\it Networks and Markets}, eds. Rauch, J. E. \& Casella A.
(Russel Sage, New York), pp. 211-257.}

\bibitem{NP}{Newman, M. \& Palmer, R. G. (1999) {\it www.arXiv.org/abs/adap-org/9908002}.}

\bibitem{Bouchaud}{Bouchaud, J.-P. (2000) {\it 
www.arXiv.org/abs/cond-mat/0008103}.}

\bibitem{JK1}
{Jain, S. \& Krishna, S. (1998)
{\it Phys. Rev. Lett.} {\bf 81}, 5684-5687.}

\bibitem{Dyson}{Dyson, F. (1985) {\it Origins of Life} (Cambridge Univ. Press
Cambridge, UK).}

\bibitem{FKP}
{Farmer, J. D., Kauffman, S. \& Packard, N. H. (1986) {\it Physica}
{\bf D22} 50-67.}

\bibitem{BFF}
{Bagley, R. J., Farmer, J. D. \& Fontana, W. (1991) 
in {\it Artificial Life II}, eds. Langton, C. G., Taylor, C., Farmer, J. D.
\& Rasmussen, S.
(Addison Wesley, Redwood City),
pp. 141-158.}

\bibitem{FB}{Fontana, W. \& Buss, L. (1994)
{\it Bull. Math. Biol.} {\bf 56}, 1-64.}

\bibitem{JK2}
{Jain, S. \& Krishna, S. (1999) 
{\it Computer Physics Comm.} {\bf 121-122}, 116-121.}

\bibitem{JK3}
{Jain, S. \& Krishna, S. (2001) 
{\it Proc. Natl. Acad. Sci. (USA)} {\bf 98}, 543-547.} 

\bibitem{JK4}
{Jain, S. \& Krishna, S. (2001) unpublished.}

\bibitem{Eigen}{Eigen, M. (1971)
{\it Naturwissenschaften} {\bf 58}, 465-523.}

\bibitem{Kauffman2}{Kauffman, S.A. (1971)
{\it J. Cybernetics} {\bf 1}, 71-96.}

\bibitem{Rossler}{Rossler, O. E. (1971)
{\it Z. Naturforschung} {\bf 26b}, 741-746.}

\bibitem{Seneta}
{Seneta, E. (1973) {\it Non-Negative Matrices} (George Allen
and Unwin, London).}

\bibitem{Paine}
{Paine, R. T. (1969) 
{\it Am. Nat.}
{\bf 103}, 91-93.}

\bibitem{Pimm}{Pimm, S. L. (1991) {\it The Balance of Nature?
Ecological Issues in the Conservation of Species and Communities} 
(Univ. of Chicago Press, Chicago).}

\bibitem{JTM}
{Jord\'{a}n, F., Tak\'{a}cks-S\'{a}nta, A., Moln\'{a}r, I. (1999)
{\it OIKOS} {\bf 86}, 453-462.}

\bibitem{SMo} {Sol\'{e}, R. V. \& Montoya, J. M. (2000) 
{\it www.arXiv.org/abs/cond-mat/0011196}.}

\bibitem{Schumpeter}{Schumpeter, J. A. (1939) {\it Business Cycles:
A theoretical, Historical and Statistical Analysis of the Capitalist
Process} (McGraw
Hill, New York).}

\bibitem{CRA}
{Cohen, M. D., Riolo, R. L. \& Axelrod, R. (2001) 
{\it Rationality and Society} {\bf 13}, 5-32.}

\bibitem{VJE}{Valentine, J. W., Jablonski, D. \& and Erwin, D. H. (1999)
{\it Development} {\bf 126}, 851-859.}

\bibitem{Erwin}{Erwin, D. H. (1996) {\it Sci. Am.} {\bf 275}, No. 1, 72-78.}

\bibitem{WS} {Watts, D.J. \& Strogatz, S. H. (1998)
{\it Nature} {\bf 393}, 440-442.}

\bibitem{BA} {Barabasi, A.-L. \& Albert, R. (1999)
{\it Science} {\bf 286}, 509-512.}

\bibitem{Watts}{Watts, D. J. (1999) {\it Small Worlds: The dynamics of Networks 
between Order and Randomness} (Princeton Univ. Press, Princeton).}

\bibitem{AB}{Albert, R. \& Barabasi, A.-L. (2001)
{\it www.arXiv.org/abs/cond-mat/0106096}.}

\end{thebibliography}
\end{document}